\documentclass[PRL,aps,twocolumn,superscriptaddress,10pt,showpacs]{revtex4-1}
\usepackage{graphicx}% Include figure files
\usepackage{epstopdf}
\usepackage{color}
\usepackage{amsmath,amssymb}
\usepackage{fixmath}
\usepackage{caption}
\usepackage{subcaption}
\usepackage{arydshln}
\usepackage{array}

\usepackage [english]{babel}
\usepackage {soul}
\usepackage [autostyle, english = american]{csquotes}
\MakeOuterQuote{"}

\begin{document}

\title{From first- to second-order phase transitions in hybrid improper ferroelectrics through entropy stabilisation}

\author{Fernando Pomiro} 
\affiliation{Department of Chemistry, University of Warwick, Gibbet Hill, Coventry, CV4 7AL, United Kingdom.}

\author{Chris Ablitt}
\affiliation{Department of Chemistry, University of Warwick, Gibbet Hill, Coventry, CV4 7AL, United Kingdom.}

\author{Nicholas C. Bristowe}
\affiliation{School of Physical Sciences, University of Kent, Canterbury, CT2 7NH, United Kingdom.}

\author{Arash A. Mostofi}
\affiliation{Departments of Materials and Physics, and the Thomas Young Centre for Theory and Simulation of Materials, Imperial College London, Exhibition Road, London SW7 2AZ, United Kingdom.}

\author{Coongjae Won}
\affiliation{Laboratory for Pohang Emergent Materials and Max Plank POSTECH Center for Complex Phase Materials, Pohang University of Science and Technology, Pohang, Korea.}

\author{Sang-Wook Cheong}
\affiliation{Laboratory for Pohang Emergent Materials and Max Plank POSTECH Center for Complex Phase Materials, Pohang University of Science and Technology, Pohang, Korea.}
\affiliation{Rutgers Center for Emergent Materials and Department of Physics and Astronomy, Rutgers University, Piscataway, New Jersey, USA.}

\author{Mark S. Senn}
\email{m.senn@warwick.ac.uk}
\affiliation{Department of Chemistry, University of Warwick, Gibbet Hill, Coventry, CV4 7AL, United Kingdom.}

\date{\today}

\begin{abstract}

Hybrid improper ferroelectrics (HIFs) have been intensely studied over the last few years to gain understanding of their temperature induced phase transitions and ferroelectric switching pathways. Here we report a switching from first- to second-order phase transition pathway for topical HIFs Ca$_{3-x}$Sr$_{x}$Ti$_2$O$_7$, which is driven by the differing entropies of the phases that we identify as being associated with the dynamic motion of octahedral tilts and rotations. A greater understanding of the transition pathways in this class of layered perovskites, which host many physical properties that are coupled to specific symmetries and octahedral rotation and tilt distortions -such as superconductivity, negative thermal expansion, fast ion conductivity, ferroelectricity, among others- is a crucial step in creating novel functional materials by design.
\end{abstract}

\pacs{75.25.Dk, 77.80.B-}

\maketitle

\section{Introduction}

Ferroelectricity in the solid state can arise due to a variety of physical phenomena, including single ion effects such as lone-pair and second-order Jahn-Teller ordering \cite{Ederer2004}. Some of the systems that exhibit these phenomena are BaTiO$_{3}$ \cite{Hippel1946} and PbTiO$_{3}$ \cite{Smolenskii1950, Roberts1950}, where ferroelectricity is driven by a single, zone-center polar lattice distortion, which gives rise to a spontaneous polarization. These kind of ferroelectrics are known as proper because the polarisation is the primary order-parameter (OP) of the phase transition from which the property evolves. In contrast, in some materials the instability of a polar mode (at the $\Gamma$ point) is a slave process to another instability, or instabilities that act as the primary OP \cite{Indenbom1960}. Such transitions include the so-called hybrid improper ferroelectrics (HIFs) \cite{Benedek2011, Benedek2015}, where two non-polar lattice distortions, e.g. octahedral rotation and tilt, couple to a polar lattice mode via a so-called trilinear coupling. This hybrid mechanism is responsible for the polar symmetry ($A2_1am$ space group) observed at room-temperature in the Ruddlesden-Popper perovskites (A$_{n+1}$B$_{n}$O$_{3n+1}$) with $n = 2$ (RP2) in low-Sr doped Ca$_{3-x}$Sr$_{x}$Ti$_{2}$O$_{7}$ (0 $\leq x \leq$ 0.9), with an experimentally verified switchable polarisation for the members with $x$=0, 0.54 and 0.85 of 8, 4.2 and 2.4~$\mu$C$\,$cm$^{-2}$, respectively \cite{Oh2015}. There has been significant interest in the unusual domain structure accessible in these materials, including head-to-head and tail-to-tail charged domain walls, and topologically protected Z$_4$ vortex domain structures \cite{Oh2015,Huang2016,Huang2016b,Smith2019}. This richness of domain structure arises from the multidimensional nature of the OP that have tilt and rotation character and span the two dimensional irreducible representations (Irreps) X$_3^-$ and X$_2^+$. Understanding how OPs evolve as a function of external stimuli, such as temperature is hence vital for controlling the properties of these materials which are intertwined with the associated domain structure. In Ca$_{3}$Ti$_{2}$O$_{7}$, rotation and tilt OPs have been shown to evolve approximately linearly with each other and polarisation between 100-500 K \cite{Senn2015}, providing evidence for the trilinear coupling mechanism. In the isostructural Ca$_{3}$Mn$_{2}$O$_{7}$, a complex competition between lattice modes of different symmetry leads to pronounced uniaxial negative thermal expansion driven by a phase coexistence between $A2_1am$ and $Acaa$ over a large temperature range \cite{Senn2015, Senn2016}, hampering any measurment of ferroeletric properties. Phase transitions in HIFs Ca$_{3-x}$Sr$_{x}$Ti$_{2}$O$_{7}$ with increasing \emph{x} and temperature have been reported from $A2_1am$ to a phase with only octahedral tilts ($P4_2/mnm$) \cite{Huang2016}, however, how the multidimensional OPs that give rise to the HIFs phase and its rich domain physics evolve remain unclear. Here we present a detailed crystallographic study, coupled with symmetry analysis and first-principles simulations, of the phase transitions taking place with temperature in the two extreme end members of the HIFs phase ($x$ = 0 and 0.85) Ca$_{3-x}$Sr$_{x}$Ti$_{2}$O$_{7}$ RP2 family. In Ca$_{3}$Ti$_{2}$O$_{7}$ we report a competing ground-state structure at very high temperature, showing a sudden strongly first-order change in the OPs that leads to a phase coexistence between $A2_1am$ and $Acaa$. In Ca$_{2.15}$Sr$_{0.85}$Ti$_{2}$O$_{7}$, a gradual change in the magnitude and direction of the OP occurs continuously with temperature spanning five distinct crystallographic symmetries from the polar $A2_1am$ structure at room temperature to the aristotype tetragonal $I4/mmm$ phase at high temperature (via $P4_2/mnm$), allowing for a second-order-like behaviour. This comparative study between the two family members, when taken with our first-principles ground-state calculations, allows us to identify that it is the enhanced entropy associated with octahedral dynamic tilts, over dynamic rotations, which serves to dictate the transition pathways in these materials. Our insights will allow for a greater control of the phase diagram of RP structures and their associated domain structures and technologically relevant properties.
 
\section{Methods}

\subsection{Sample Preparation}
High-quality polycrystalline samples of Ca$_{3}$Ti$_{2}$O$_{7}$ and Ca$_{2.15}$Sr$_{0.85}$Ti$_{2}$O$_{7}$ were prepared using the standard solid-state synthesis method at 1300-1500$^{o}$C. Stoichiometric amounts of CaCO$_{3}$ (Alfa Aesar 99.95\%), SrCO$_{3}$ (Alfa Aesar 99.99\%) and TiO$_{2}$ (Alfa Aesar Puratronic 99.995\%) powders were well mixed, ground, pelletized and then heated for 30~h in air. 

\subsection{Data collection and data analysis.}
High-resolution synchrotron powder diffraction experiments were performed in Debye-Scherrer geometry at I11, Diamond Light Source, Didcot, UK. The samples were packed into a quartz capillary with a diameter of 0.3~mm. The experiments were performed on warming and on cooling in the temperature range from 300 to 1250~K using the MYTHEN detector (Position Sensitive Detector or PSD) measuring a pattern every 4~K. In Ca$_{2.15}$Sr$_{0.85}$Ti$_{2}$O$_{7}$ we have also performed measurements using the Multi-Analyser Crystal detectors (MAC) at 300, 473, 623, 773 and 873~K with the aim of obtaining diffraction patterns with a much lower background and a resolution one order of magnitude higher than PSD. The zero-point error, wavelength ($\approx$ 0.825~\AA) and instrument contribution to the peak profile were determined against a NIST 640 Si standard and fixed in all subsequent analysis. An empirical absorption correction was applied to the fitted Rietveld model based on cylindrical corrections \cite{Sabine1998}. Temperature control was achieved with a hot-air-blower from 300~K to 1250~K. The collected patterns were analysed with the Rietveld method using TOPAS with the JEDIT interface \cite{Coelho2009,Evans2010}.  

For the refinements against the 300 K data, random starting values for the distortion modes were generated between -0.04 and 0.04 (except for the mode a3 that correspond with the Irrep [0,0,0] $\Gamma_{5}^{-}$(a;-a)[Ca1:b:dsp]Eu(a) which was used to fix the origin in the x direction), and isotropic thermal parameters were refined for all sites. Refinements with new initial starting values were repeated multiple times, but no false minima were observed. Subsequent refinements at different temperatures were performed sequentially using the previous refinement output as the new input. A small amount of anisotropic peak broadening in the diffraction pattern was modelled using the Stephens phenomenological description considering the structure as pseudo tetragonal \cite{Stephens1999}.

The displacive modes are characterized by the point of the Brillouin zone of the parent structure it belongs to, the position in reciprocal space as well as the short hand symbol is given e.g. [1/2,1/2,0]X$_3^-$. The following number and sign e.g. "3-" correspond to Irreps as tabulated by Stokes and Hatch \cite{Stokes1989}. The next label which follows is the branching of the distortion mode (related to the order parameter directions). For example (a;0) indicates that the order parameter is active in one of the doubly degenerate directions whereas (a;a) would imply that it is active in both directions and in this case the child structure would retain tetragonal symmetry. Next, the site label and Wyckoff site symmetry is given (with respect to the parent $I4/mmm$ structure, e.g. [Sr1:b]) on which the distortion mode acts. Finally the irreducible representation of the distortion with respect to the Wyckoff site point group symmetry is given (e.g. Eu for point group D4h (4/mmm)-Wyckoff site 2b).

In Ca$_{2.15}$Sr$_{0.85}$Ti$_{2}$O$_{7}$ we performed variable temperature refinements against the collected data with the highest symmetry subgroup $P2_1nm$ that is common to $P4_2/mnm$, $Amam$, $Pnnm$ and $A2_1am$ models. We used the web based ISODISTORT tool \cite{Campbell2006} to generate this child structure parametrized in terms of distortion modes which may be refined directly in the Rietveld refinement program Topas (see Supplemental Material \cite{SI}). $P2_1nm$ (number 31) basis=(1,-1,0),(1,1,0),(0,0,1), origin=(0,0,0) was generated from the parent $I4/mmm$ structure of Sr$_3$Ti$_2$O$_7$ published by Elcombe \cite{Elcombe1991} in which the Ti atom sits on the (0, 0, z) site using ISODISTORT (see .str file in the Supplemental Material \cite{SI}). This model has 41 internal degrees of freedom (see Table~\ref{tab:phasesDoF}) and with the proper symmetry constraints we can reproduce the $P4_2/mnm$, $Amam$, $Pnnm$ and $A2_1am$ models. Initially we perform a completely unbiased search for the active modes at all temperatures against our variable temperature PSD data. The result of this search is shown in Fig.~S4 in the Supplemental Material \cite{SI}.  

\subsection{First-principles simulations.}

First-principles simulations were performed using CASTEP v.7.0.3 \cite{CASTEP}. These calculations employed the PBEsol \cite{PBEsol} functional to approximate exchange and correlation effects, a 1400~eV plane-wave cut-off energy with a grid twice as dense for the electron density and a $7 \times 7 \times 1$ Monkhorst--Pack k-point grid for sampling the first Brillouin zone of the $I4/mmm$ phase of Ca$_{3-x}$Sr$_x$Ti$_2$O$_7$. This grid was scaled appropriately for other structures. Structural relaxations used a force tolerance of 0.5~meV/{\AA} and a stress tolerance of 10~MPa. The nuclei and core electrons of all ions were represented using norm-conserving pseudopotentials (details are given in Table~S1 in the Supplemental Material \cite{SI}) and the virtual crystal approximation (VCA) was used to simulate a solid solution between Ca and Sr ions. We studied eight competing RP2 phases (the first eight listed in Table \ref{tab:phasesDoF}). Initial structures were generated by freezing-in small distortions to a relaxed $I4/mmm$ structure using ISODISTORT and then letting the cell and internal coordinates relax. $Pnnm$ is not shown in Fig.~\ref{Fig_3} since for all compositions these structures relaxed to the tetragonal $P4_2/mnm$ phase.

We used the nudged elastic band (NEB) method \cite{Henkelman2000NEB} as implemented in ASE \cite{Hjorth2017ase} and adapted for use with CASTEP \cite{CASTEPNEBwrapper} to compute the minimum energy pathway between different phases. Small modifications were made to the NEB wrapper to integrate with simulations employing the VCA. Initial transition pathways were found by linear interpolation to give five intermediate trajectories (excluding the initial and final relaxed phases). The NEB method was then performed iteratively until the energies of intermediary structures converged to within 0.1 meV/atom. We used FINDSYM \cite{Findsym} with a tolerance of 0.001 to detect the space group symmetry of all structures and AMPLIMODES \cite{Amplimodes1, Amplimodes2} to compute mode amplitudes relative to the relaxed $I4/mmm$ parent. It was found that the amplitude value changed depending on the symmetry assignment of the distorted phase and therefore we used ISODISTORT to reduce the symmetry of all structures to $P2_1$ (the highest-symmetry subgroup common to all structures) by ``freezing in'' zero-amplitude distortions where necessary. All NEB iterations are plotted in Figs.~S8b and S8c in the Supplemental Material \cite{SI} to illustrate the extent that the minimum energy pathway deviated from the initial interpolation. The implementation of the NEB method required fixed cell boundary conditions and therefore the tetragonal unit cell of the relaxed $I4/mmm$ parent was imposed on all NEB structures (with end-point phases relaxed with this cell fixed). Since this method in its current implementation requires fixed strain end-points for the pathways, and the $Acaa$ and $A2_1am$ phases have substantially different lattice parameters, a fact that is intrinsically linked to the unusual thermal expansion properties of the $Acaa$ and the RP1 related phases \cite{Senn2015,Senn2016,Ablitt2018,Ablitt2017}, an additional set of calculations is performed in which a linear interpolation between the optimised trajectories of the NEB are frozen-in and lattice parameters relaxed.

\begin{figure}[h!]
\includegraphics[width=\columnwidth] {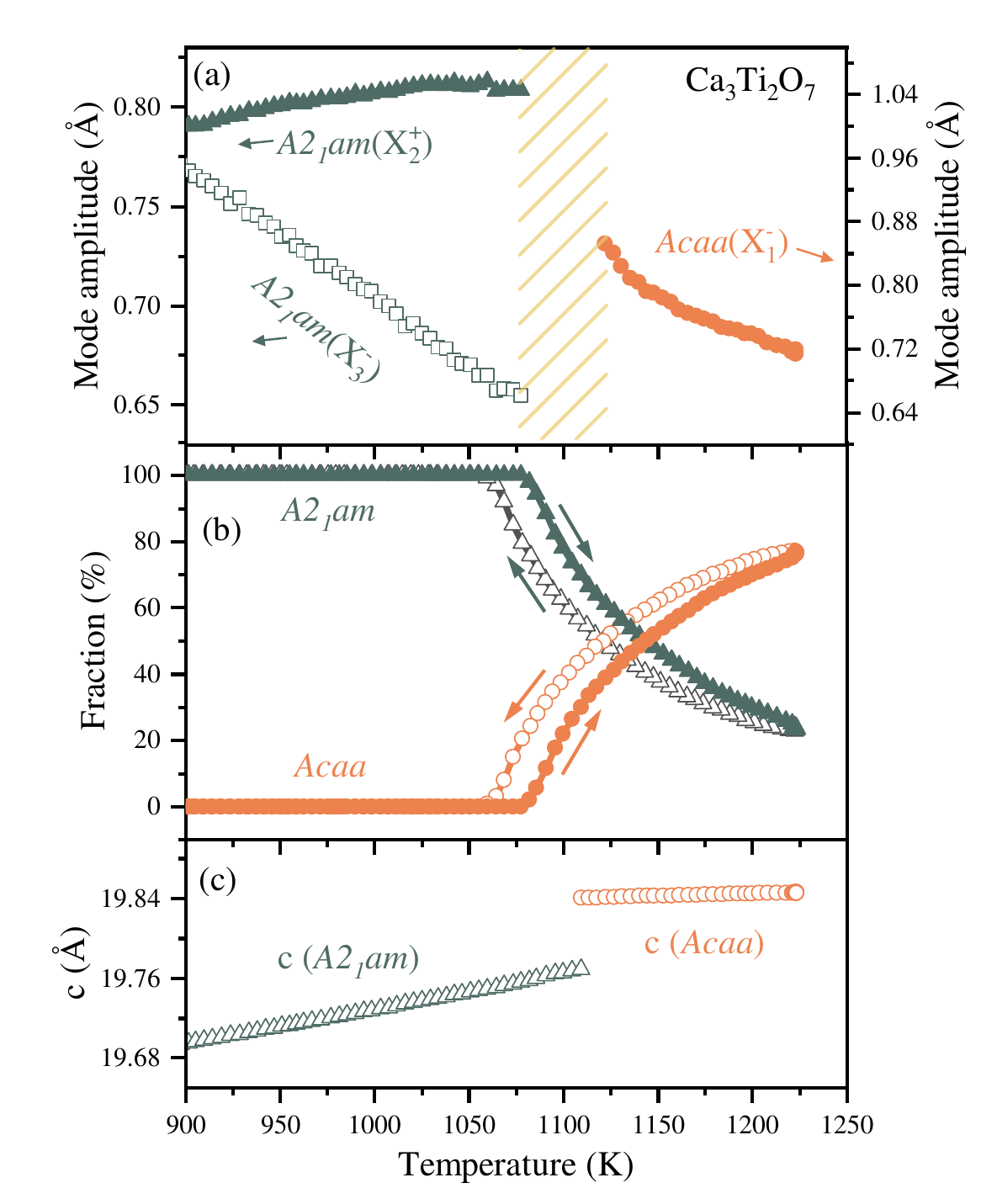}
\caption{\label{Fig_1} Thermal evolution of (a) distortion modes of symmetry $X_2^+$, $X_3^-$ and $X_1^-$, (b) phase fraction in Ca$_{3}$Ti$_{2}$O$_{7}$ and (c) \textit{c} lattice parameter in Ca$_{3}$Ti$_{2}$O$_{7}$.}
\end{figure}

\section{Results}

\begin{figure}[h!]
\includegraphics[width=\columnwidth] {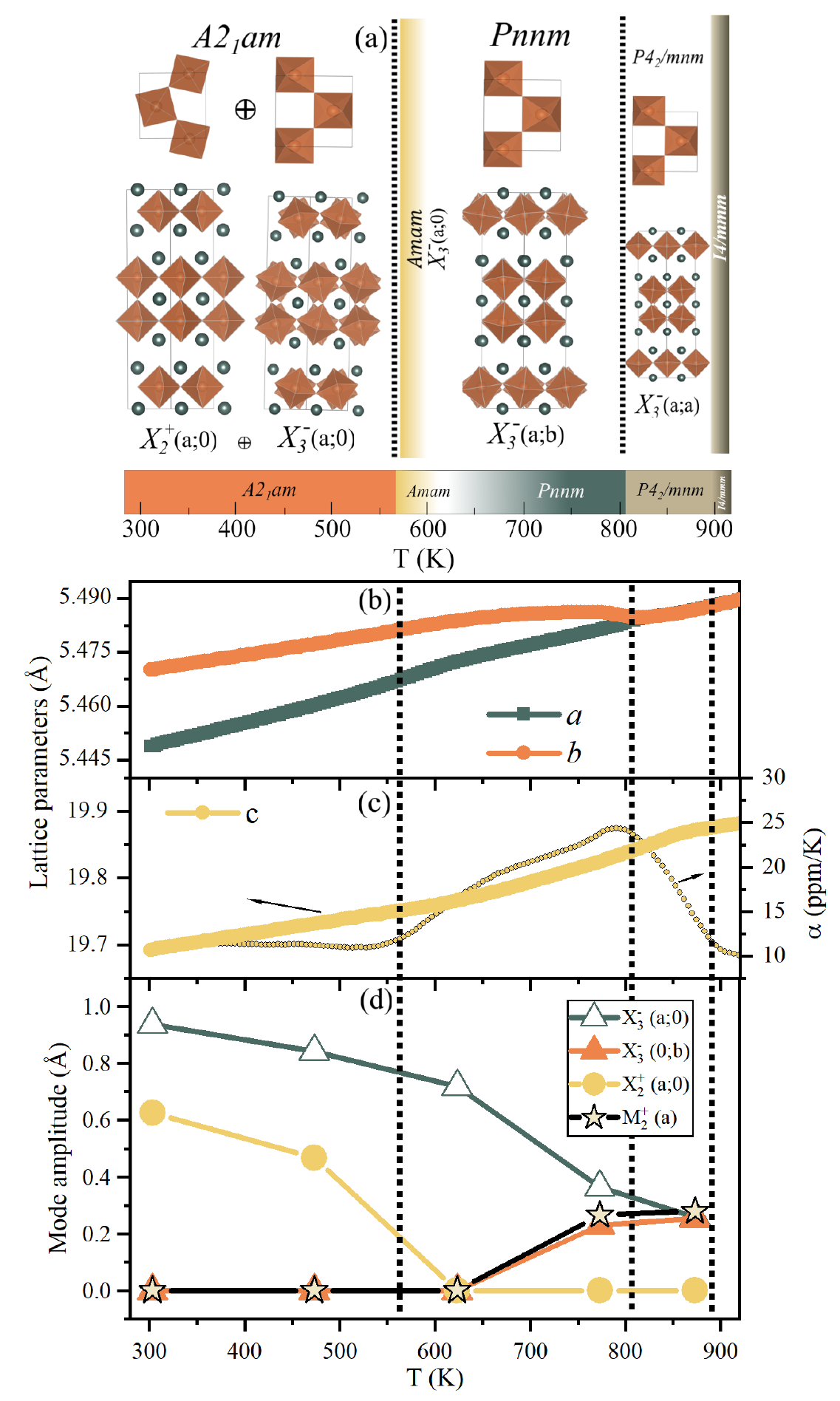}
\caption{\label{Fig_2} (a) Thermal evolution of the rotations and tilts of the BO$_6$ octahedral together with the space group and relevant symmetry modes for each phase in Ca$_{2.15}$Sr$_{0.85}$Ti$_{2}$O$_{7}$. (b, c) Thermal evolution of the lattice parameters \emph{a}, \emph{b} and \emph{c} and the thermal expansion coefficient of \emph{c}. (d) Thermal evolution of the distortion modes of character $X_3^-$ (a;0), $X_3^-$(0,b), $X_2^+$(a;0) and M$_{2}^{+}$.}
\end{figure}

Temperature dependent evolution of the distortion modes, phase fraction and $c$ lattice parameter in Ca$_{3}$Ti$_{2}$O$_{7}$ are summarised in Fig.~\ref{Fig_1}. Inspection of the diffraction patterns at 300~K of Ca$_{3}$Ti$_{2}$O$_{7}$ and Ca$_{2.15}$Sr$_{0.85}$Ti$_{2}$O$_{7}$ revealed that, whilst the main reflections can be indexed with the tetragonal $I4/mmm$ aristotype structure, additional reflections at (\emph{h}+$\frac{1}{2}$, \emph{k}+$\frac{1}{2}$, \emph{l}) were present, together with the orthorhombic splitting of basic \textit{hhl} reflections in the aristotype tetragonal setting. These superstructure reflections index on the previously proposed $A2_1am$ model \cite{Huang2016,Senn2015}. The $A2_1am$ structure (a$^{-}$a$^{-}$c$^{+}$/a$^{-}$a$^{-}$c$^{+}$ in Glazer notation) corresponds to the direct sum of the symmetry spaces spanned by $Acam$ (a$^{0}$a$^{0}$c$^{+}$/a$^{0}$a$^{0}$c$^{+}$) and $Amam$ (a$^{-}$a$^{-}$c$^{0}$/a$^{-}$a$^{-}$c$^{0}$) and is due to the simultaneous ordering of $X_2^+$(a;0) (in-phase rotation of the BO$_{6}$ octahedra about the $c$-axis) and $X_3^-$(a;0) (out-of-phase tilting of the BO$_{6}$ octahedra in the $a-b$ plane) Irreps of the parent space group $I4/mmm$ (a$^{0}$a$^{0}$c$^{0}$/a$^{0}$a$^{0}$c$^{0}$) under the ordering propagation vector [$\frac{1}{2}$,$\frac{1}{2}$,0] (see Fig.~\ref{Fig_2}).  

In our previous work \cite{Senn2015}, we demonstrated a hybrid coupling between the $X_2^+$(a;0) and $X_3^-$(a;0) OPs in Ca$_{3}$Ti$_{2}$O$_{7}$ between 100 and 500~K without observing any phase transition in this temperature range. In this work, our aim is to understand what happens in the structure at higher temperatures and how the phase transition from the ferroelectric orthorhombic $A2_1am$ to the paraelectric phase proceeds since this will likely give insight into the ferroelectric switching pathway. Fig.~\ref{Fig_1} summarises the obtained results for Ca$_{3}$Ti$_{2}$O$_{7}$. There are no discontinuities in mode or lattice parameters from 300 to 1100~K. However, Fig.~\ref{Fig_1}b shows phase coexistence in the sample above 1100~K, with a high temperature phase whose lattice parameters deviate significantly from those for the low temperature phase  (see Fig.~\ref{Fig_1}c), whilst still being orthorhombic. At high temperatures an $Acaa$ model with a single out-of-phase octahedral rotation [$X_1^-$(a;0)] provides the best fit to the data (see the Supplemental Material \cite{SI} for full details by which this assignment is made). This phase transition pathway means that there must be a strong discontinuity on going from in-phase $X_2^+$(a;0) to out-of-phase $X_1^-$(a;0) octahedral rotations (see Fig.~\ref{Fig_1}a), as it is hard to envisage a scenario in which these could both coexist in the same structure in a physically meaningful way. Such a structural transformation should therefore occur discontinuously with a first-order character. Symmetry analysis confirms that there is no direct pathway along which the $A2_1am$ phase may distort to reach its high temperature form of $Acaa$. The first-order character of this phase transition is experimentally confirmed by our variable temperature synchrotron X-ray diffraction data, showing a clear coexistence of the $A2_1am$ and $Acaa$ phases with an hysteresis of about 18~K on heating and cooling across the phase transition (see Fig.~\ref{Fig_1}b and Fig.~S3 in the Supplemental Material \cite{SI}). Fig.~\ref{Fig_1}c displays the thermal evolution of the \textit{c} lattice parameters (see Supplemental Material \cite{SI}), showing a clear discontinuity around the phase transition temperature and a decrease in the thermal expansion coefficient echoing of the negative thermal expansion that has been shown to be specific to the Ca$_{3-x}$Sr$_{x}$Mn$_{2}$O$_{7}$ ($x = 0-1.5$) $Acaa$ phase at lower temperatures \cite{Senn2015,Senn2016}. Our results also explain the origin of the previously reported first-order phase transition at around 1100~K from differential scanning calorimetry, with an endothermic peak at 1100~K during the heating cycle and an exothermic peak at 1082~K during the cooling cycle (hysteresis of 18~K) \cite{Liu2015}.                                     

The structural phase transitions observed with temperature in Ca$_{2.15}$Sr$_{0.85}$Ti$_{2}$O$_{7}$ have a very different behaviour to those observed in Ca$_{3}$Ti$_{2}$O$_{7}$. The first big difference is in the thermal evolution of the lattice parameters, which show a gradual change with temperature reaching a tetragonal phase around 800~K (see Fig.~\ref{Fig_2}b). In Fig.~\ref{Fig_2}c we show the thermal expansion coefficient ($\alpha$) of the $c$ lattice parameter, which is the derivative of the strain with respect to temperature. Four regimes are evident, two metrically orthorhombic in the ranges 300-550~K and 550-800~K and two metrically tetragonal, the first one in the range 800-900~K and the other at $T \geqslant$ 900~K. These four regimes provide the first evidence of the phase transitions taking place in our sample. In Fig.~\ref{Fig_2}d we show the evolution of the distortion modes obtained through refinement of the high-resolution diffraction data with the space group that we have assigned on the basis of the analysis discussed in the methods section. From this analysis the four distinct regimes that we observe are $A2_1am$, $(Amam)Pnnm$, $P4_2/mnm$ and $I4/mmm$ (see Fig.~\ref{Fig_2}a). These correspond to continuous decrease in amplitude of the $X_2^+$ OP, followed by a gradual change in magnitude and direction of the $X_3^-$ OP and finally reaching a zero value of the amplitude of both OPs, $X_2^+$ and $X_3^-$. This pathway traces out the sequence of phase transitions on warming $A2_1am$ $\rightarrow$ $(Amam)$ $\rightarrow$ $Pnnm$ $\rightarrow$ $P4_2/mnm$ $\rightarrow$ $I4/mmm$. This continuous sequence provides an explanation for the second-order nature of the phase transitions, which is consistent with the OP amplitude refined against the diffraction data, the continuous evolution of lattice parameters, and the absence of any phase coexistence. This is a much richer phase diagram than that previously proposed \cite{Kratochvilova2019}, which assumed a first-order phase transition from $A2_1am$ directly to $P4_2/mnm$ at this composition. Thus in summary, $x = 0$ shows first-order while $x = 0.85$ shows second-order phase transitions.
        
\begin{figure*}
\centering
\includegraphics[width=0.8\textwidth] {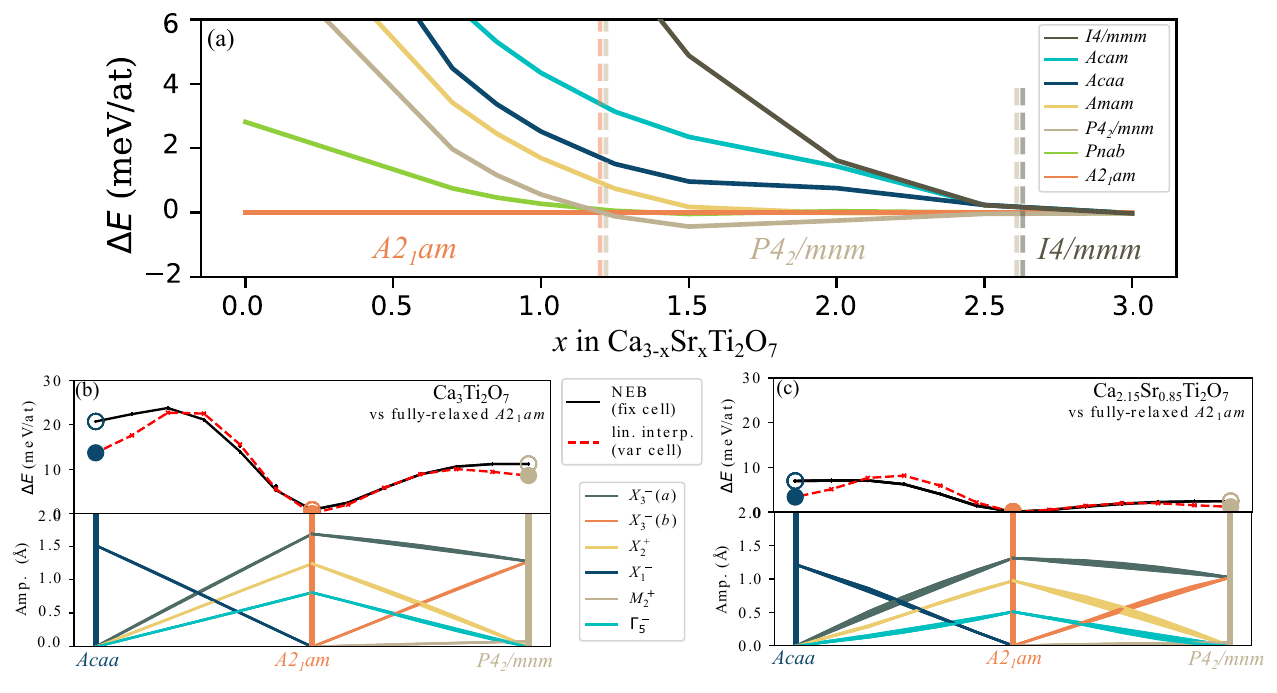}
\caption{\label{Fig_3} DFT simulation results: (a) energy difference of several Ruddlesden--Popper phases compared to the polar $A2_1am$ phase for Ca$_{3-x}$Sr$_x$Ti$_2$O$_7$ compounds; (b) and (c) transition pathways between the $A2_1am$ ground-state and $Acaa$ (left) and $P4_2/mnm$ (right) phases for Ca$_3$Ti$_2$O$_7$ and Ca$_{2.15}$Sr$_{0.85}$Ti$_2$O$_7$, respectively. The top panels of (b) and (c) show phase energies (relative to fully-relaxed $A2_1am$) for all transitions. The end-point phases for fixed cell boundary conditions (see Methods section) are shown as coloured open circles with black crosses connected by black lines showing the intermediate trajectories. The pathways for relaxed lattice parameters (see Methods section) are displayed as red dashed lines connecting filled circles representing phases with relaxed cells. The bottom panel display the mode amplitudes, labelled according to their Irrep in the parent $I4/mmm$ phase, for one-step NEB pathways.}
\end{figure*}

\begin{table}
  \begin{tabular}{ccc}
    \textbf{Space Group} & \textbf{irreps (rel. $I4/mmm$)} & \textbf{DoF} \\[2mm]
    $I4/mmm$ (139) & - & 6 \\[1mm]
    $A2_1am$ (36) & $X_2^+$(a;0) $\oplus$ $X_3^-$(a;0) $\oplus$ $\Gamma_5^-$ & 22 \\[1mm]
    $Amam$ (63) & $X_3^-$(a;0) & 13 \\[1mm]
    $Pnnm$ (58) & $X_3^-$(a;b) $\oplus$ $M_2^+$ & 20 \\[1mm]
    $P4_2/mnm$ (132) & $X_3^-$(a;a) $\oplus$ $M_2^+$ & 13 \\[1mm]
    $Acam$ (68) & $X_2^+$(a;0) & 9 \\[1mm]
    $Acaa$ (68) & $X_1^-$(a;0) & 9 \\[1mm]
    $Pnab$ (60) & $X_1^-$(a;0) $\oplus$ $X_3^-$(a;0) & 15 \\[2mm]
    \cdashline{1-3} \\[-2mm]
    $P2_1nm$ (31) & $X_2^+$(a;0) $\oplus$ $X_3^-$(a;b) \dots & 41 \\[1mm]
    $P2_1$ (4) & $X_1^-$(a;0) $\oplus$ $X_2^+$(a;0) $\oplus$ $X_3^-$(a;b) \dots & 75 \\[1mm]
  \end{tabular}
  \caption{List of A$_3$B$_2$O$_7$ Ruddlesden--Popper space groups considered in this study. For each phase, key Irreps are given relative to the aristotype $I4/mmm$ phase, as are the number of DoF allowed in the structure (including $\Gamma_1^+$ distortions and cell strains). The first eight phases were relaxed using DFT whereas the latter two ($P2_1nm$ and $P2_1$) describe the symmetry of transition pathways encountered in nudged elastic band (NEB) simulations.}
  \label{tab:phasesDoF}
\end{table}

To understand the reason behind this switch from first- to second-order-like transitions we perform first-principles simulations. We first investigate how the energy landscape evolves with changing Ca:Sr ratio. Fig.~\ref{Fig_3}a compares the per atom energy of different phases relative to $A2_1am$. Our density-functional theory (DFT) calculations predict that all phases for Sr$_3$Ti$_2$O$_7$ relax to the $I4/mmm$ parent structure. For all lower values of $x$, phases with only an octahedral rotation about the layering axis ($Acam$ and $Acaa$) were higher in energy than any phase with octahedral tilts about an in-plane axis. For $x < 1.25$ the polar $A2_1am$ phase was the ground-state, yet for an intermediate range of $x$ ($1.25 \leq x \leq 2.5$) tetragonal $P4_2/mnm$ structures were lowest in energy. This $A2_1am \rightarrow P4_2/mnm \rightarrow I4/mmm$ sequence of ground-state phases with increasing $x$ qualitatively agrees with the room temperature experimental phase diagram \cite{Huang2016}, albeit with simulated transition boundaries shifted to higher $x$, presumably reflecting the entropic contributions in the experimental phase diagram. Our ground-state calculations are zero temperature calculations and, therefore, do not account for vibrational entropy, $S_{\mathrm{vib}}$, or zero-point energy contributions. However, in broad terms, higher vibrational entropy is linked with fewer crystallographic degrees of freedom (DoF). Noting that this implies that $S_\mathrm{vib}(I4/mmm) > S_\mathrm{vib}(P4_2/mnm) > S_\mathrm{vib}(A2_1am)$ (since the structural DoF of these phases are 6, 13 and 22, respectively -- see Table \ref{tab:phasesDoF}), it becomes clear that a shift of the simulated phase diagram to higher $x$ is actually expected when not accounting for entropy.

In order to highlight the role of entropy in the phase transition taking place in Ca$_3$Ti$_2$O$_7$ and Ca$_{2.15}$Sr$_{0.85}$Ti$_2$O$_7$, we simulated the energetics of various possible transition pathways using DFT and the nudged elastic band (NEB) method \cite{Henkelman2000NEB}. Figs.~\ref{Fig_3}b and \ref{Fig_3}c show pathways connecting the $A2_1am$ phase (centre) to the $Acaa$ (left) and $P4_2/mnm$ (right) phases for Ca$_3$Ti$_2$O$_7$ and Ca$_{2.15}$Sr$_{0.85}$Ti$_2$O$_7$, respectively. Both the end-point phases for fixed (open circles) and relaxed (filled circles) cell boundary conditions are considered (see Methods section). Since the pathways for relaxed lattice parameters are lowest in energy (filled circles in top panel of Figs.~\ref{Fig_3}b and~\ref{Fig_3}c), they will be discussed exclusively in the following section, and the significance of allowing the cell to relax is particularly evident in the decrease in energy of the $Acaa$ relaxed structures.

First it is interesting to note that the effect of the substitution from $x = 0$ to $x = 0.85$ does not substantially favour any energy minimum over another but acts more to scale the whole energy trajectory ($\Delta E_{(Acaa)} / \Delta E_{(P4_2/mnm)}$ with relaxed cell is 1.60 and 2.80 for $x = 0$ and $x = 0.85$, respectively). On the other hand, the much smaller energy scales associated with the DFT ground-state energies in $x = 0.85$ versus $x = 0$, imply that a smaller entropic contribution to the Gibbs free energy is required to affect the phase transition in the former, as observed experimentally. 

\begin{figure}[h!]
\includegraphics[width=\columnwidth] {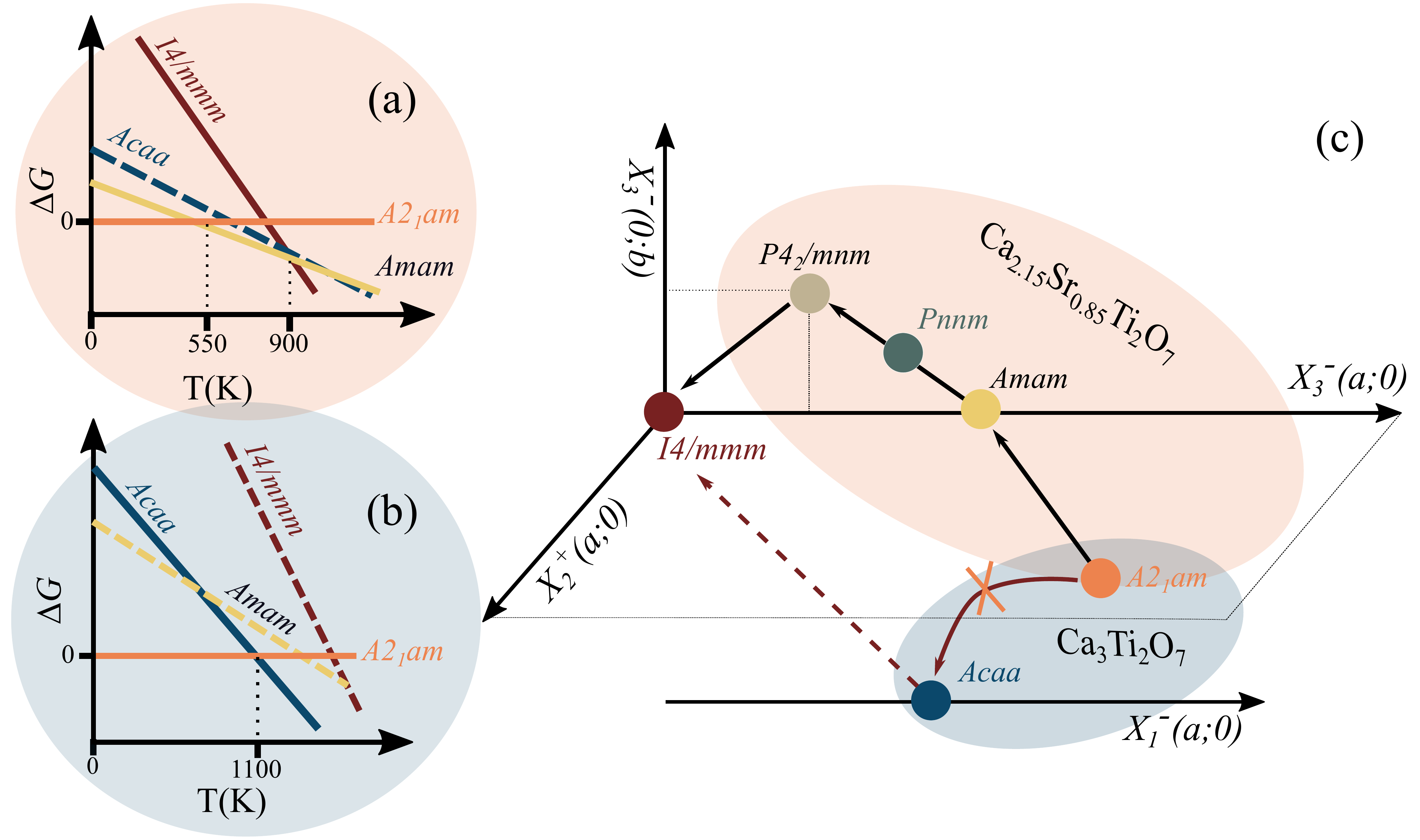}
\caption{\label{Fig_4} (a) and (b) Illustrative demonstration of how a crossover from a first- to a second-order phase transition could occur. (c) Evolution of the order parameter amplitude and direction across the phase transitions for Ca$_{3-x}$Sr$_{x}$Ti$_2$O$_7$ (x = 0 and 0.85).}
\end{figure}

DFT predicts the two-step transition pathway $A2_1am$ $\rightarrow$ $Amam$ $\rightarrow$ $Pnnm$ $\rightarrow$ $P4_2/mnm$ to have a higher activation barrier than a direct $A2_1am \rightarrow [P2_1nm] \rightarrow P4_2/mnm$ transition, although only by 1.31~meV/atom in Ca$_{2.15}$Sr$_{0.85}$Ti$_2$O$_7$ (further details are given at Supplemental Material \cite{SI}).  This is, however, in good agreement with the continuous second-order-like transformation through intermediate $Amam$ and $Pnnm$ phases that we observe experimentally in Ca$_{2.15}$Sr$_{0.85}$Ti$_2$O$_7$ at finite temperatures where the almost negligible difference in transition pathway energy will be overcome by the contribution from the higher vibrational entropy associated with higher symmetry and fewer crystallographic DoF (see Table \ref{tab:phasesDoF}).

As in both compositions the $P4_2/mnm$ phase is lower in energy, our ground-state calculations fall short to explain why Ca$_3$Ti$_2$O$_7$ transforms to $Acaa$. However, the profile of our pathway calculated for $A2_1am \rightarrow Acaa$ is consistent with the strongly first-order nature of this phase transition evident in our diffraction data. The activation barrier is over 20~meV/atom above the $A2_1am$ phase and 10~meV/atom above $Acaa$ (see Fig.~\ref{Fig_3}b top panel). The failure of our ground-state calculations to predict the correct pathway for x = 0, is interesting in itself, and implies that the entropic contributions to the Gibbs free energy are substantially greater in the $Acaa$ phase than in $P4_2/mnm$ (or $Amam$). The disagreement between our NEB calculations and experiment hence underline the interplay between the ground-state energy surface and the entropy. Relatively flat energy landscapes ($x = 0.85$), mean that entropically the transition pathway is navigated close to the saddle points of the DFT energy surface. However, for energy surfaces with deeper minima, where transition temperatures are much higher ($x = 0$), entropic contributions may end up changing global minimum and hence not just the transition pathway taken, but also the thermodynamically favoured phase. Evidence to support this hypothesis may be taken by considering Table \ref{tab:phasesDoF}, $Acaa$ (9 DoF) versus $P4_2/mnm$ or $Amam$ (13 DoF), where, as we mention before, a high number of crystallographic DoF implies more order and therefore less entropy. On a more microscopic level, these phases differ in that $Acaa$ has static rotations along, and dynamic tilts away from, the $c$-axis, while the converse is true in $P4_2/mnm$ (and $Amam$). With the benefit of hindsight it is easy to appreciate why the entropy associated with tilts (that are dynamic in the $Acaa$ phase) is higher than that associated with rotations (which are dynamic in the $Amam$ and $P4_2/mnm$ phases). The layering of the RP structure along the $c$-axis provides twice as many extra DoF for the tilts than the rotations (see Fig.~S11 at Supplemental Material \cite{SI}). These extra DoF, relieve the constraint requiring $BO_6$ octahedra to counter-tilt in neighbouring blocks such as is required for the rotations within the layers or indeed any such motions in $ABO_3$ perovskites. Alternatively, this fact may be appreciated by viewing the phonon dispersion curves of a RP1 ($n = 1$) perovskite that have an additional line with rigid unit mode character associated with tilts compared to the rotations (see Fig.~S12 at Supplemental Material \cite{SI}). However, this introduces a further puzzle: if the entropic contributions of dynamic tilts in $Acaa$ is always greater than that of the dynamic rotations in $Amam$, why does $x = 0.85$ favour this latter phase? The origin of this apparent contraction may be resolved by considering Fig.~\ref{Fig_4}, where we illustrate $\Delta G = \Delta H - T\Delta S$ (where $\Delta G$ is the Gibbs free energy, $\Delta H$ the enthalpy and $\Delta S$ the entropy of the system) for the different phases, where the $T = 0$ intercept could be considered to be equal to the ground-state energies calculated for our present compositions, which are much greater in $x = 0$ than 0.85. We then fix the slope ($\Delta S$) of the $Acaa$ for $x = 0$ and $Amam$ for $x = 0.85$ based on the experimentally observed transition temperatures. If the remaining $Acaa$, $x = 0$ and $Amam$, $x = 0.85$ lines are drawn according to their DFT ground-state energies and such that $\Delta S_{Acaa} > \Delta S_{Amam}$, consistent with tilts having more entropy than rotations, then the observed compositional dependent switching from first- to second-order phase transitions would clearly be observed. We emphasise that these plots are only illustrative, but clearly underline the significance of entropy in determining phase transition pathways in RP compounds compared to $ABO_3$ perovskites where there is not a distinction between tilts and rotations.  

Recently the importance of rotational entropy associated with the molecular methylammonium cations in metal organic perovskites in determining phase stability has been highlighted \cite{Chen2016,Wei2018}. Conceptually these ideas are related to the explanation we have given here, and our ability to tune the transition pathways in the RP structures through solid solution chemistry will make them an exciting playground for investigate of this interplay between enthalpy and entropy. A detailed quantitative understanding of the entropy difference between rotations and tilts is beyond the scope of the present work. However, a greater understanding of this will allow for a more precise control of the complex phase diagrams of RP phases and their functional properties.

\section{Conclusion}
In summary, we have shown that HIFs Ca$_{3-x}$Sr$_{x}$Ti$_2$O$_7$ have a pronounced first-order phase transition from ferroeletric $A2_1am$ to $Acaa$ at $x = 0$ but this switches to a continuous second-order phase transition by $x = 0.85$. This second-order phase transition to $P4_2/mnm$ is possible via a continuous decrease in magnitude of the $X_2^+$ (rotation) order parameter followed by a rotation and decrease in magnitude of $X_3^-$ (tilt) order parameter direction. As the order parameter direction rotates it passes through phases of specific symmetry as evident by our detailed crystallographic analysis and summaries in Fig.~\ref{Fig_4}c. Our DFT ground-state calculations of the possible transition pathways in these materials confirm this picture of a first-order phase transition at $x = 0$ and a second-order at $x = 0.85$. Furthermore, they serve to highlight the importance of entropy associated with the different phase in which either octahedral rotations ($Amam$) or tilts ($Acaa$) remain dynamic, in selecting out either first or second-order transition pathways in this system. We suggest that a broader understanding of the differing entropies associated with tilts and rotations will help rationalise the phase diagram of the technologically important RP phases.

\begin{acknowledgments}
This work was supported by ESPRC grant no. EP/S027106/1. The synchrotron beam time used in this paper was at I11 through the Diamond Light Source Block Allocation Group award “Oxford/Warwick Solid State Chemistry BAG to probe composition-structure-property relationships in solids” (EE18786). The work at Rutgers University was supported by the DOE under Grant No. DOE: DE-FG02-07ER46382, and the work at Postech was supported by the National Research Foundation of Korea (NRF) funded by the Ministry of Science and ICT(No. 2016K1A4A4A01922028). We are grateful to the UK Materials and Molecular Modelling Hub for computational resources, which is partially funded by EPSRC (EP/P020194/1). FP would like to acknowledge the IAS at University of Warwick and the Marie Skłodowska-Curie Actions for the WIRL-COFUND fellowship. CA was supported through a studentship in the Centre for Doctoral Training on Theory and Simulation of Materials at Imperial College London funded by the EPSRC (EP/L015579/1). AAM acknowledges the support of the Thomas Young Centre through grant TYC-101. MSS acknowledges the Royal Society for a University Research Fellowship (UF160265).
\end{acknowledgments}

\bibliography{draft_02_01_2020_bib}

\end{document}